\documentclass[aps,pra,twocolumn,superscriptaddress,showpacs,floatfix]{revtex4}
\usepackage{graphicx,dcolumn,bm,hyperref,amsmath,amssymb,xspace,epsfig,float,array,multirow,amsfonts}
\usepackage[vcentermath]{youngtab}
\usepackage{braket}

\usepackage{xcolor} 

\newcommand{\aho}{a_\mathrm{ho}}
\newcommand{\gd}{g_{1\mathrm{D}}}
\newcommand{\EF}{E_\mathrm{F}^{(\infty)}}
\newcommand{\kF}{k_\mathrm{F}}
\newcommand{\kB}{k_\mathrm{B}}
\newcommand{\LdB}{\Lambda_\mathrm{dB}}

\begin{document}

\title{High-momentum tails as magnetic structure probes \\ for strongly-correlated $SU(\kappa)$ fermionic mixtures in one-dimensional traps}

\author{Jean Decamp}
\affiliation{Universit\'e C\^ote d'Azur, CNRS, INLN, France}

\author{Johannes J\"unemann}
\affiliation{Johannes Gutenberg-Universit\"at, Institut f\"ur Physik, Staudingerweg 7, 55099 Mainz, Germany}
\affiliation{MAINZ - Graduate School Materials Science in Mainz, Staudingerweg 9, 55099 Mainz, Germany}

\author{Mathias Albert}
\affiliation{Universit\'e C\^ote d'Azur, CNRS, INLN, France}

\author{Matteo Rizzi}
\affiliation{Johannes Gutenberg-Universit\"at, Institut f\"ur Physik, Staudingerweg 7, 55099 Mainz, Germany}

\author{Anna Minguzzi}
\affiliation{Universit\'e Grenoble-Alpes, CNRS, LPMMC, BP166, F-38042 Grenoble, France}

\author{Patrizia Vignolo}
\affiliation{Universit\'e C\^ote d'Azur, CNRS, INLN, France}

\date{\today}

\begin{abstract}
A universal $k^{-4}$ decay of the large-momentum tails of the momentum distribution, fixed by Tan's contact coefficients, constitutes a direct signature of strong correlations in a short-range interacting quantum gas. Here we consider a repulsive multicomponent Fermi gas under harmonic confinement, as in the experiment of Pagano et al. [Nat. Phys. {\bf 10}, 198 (2014)], realizing a gas with tunable $SU(\kappa)$ symmetry. We  exploit an exact solution at infinite repulsion to  show a direct correspondence between the value of the Tan's contact for each of the $\kappa$ components of the gas and the  Young tableaux for the $S_N$ permutation symmetry group identifying  the magnetic structure of the ground-state. This opens a route for the experimental determination of magnetic configurations in cold atomic gases, employing only standard (spin-resolved) time-of-flight techniques. Combining the exact result with matrix-product-states simulations, we obtain the Tan's contact at all values of  repulsive interactions. We show that a local density approximation (LDA) on the Bethe-Ansatz equation of state for the homogeneous mixture is in excellent agreement  with the results for the harmonically confined gas. At strong interactions, the LDA predicts a scaling behavior of the Tan's contact. This provides a useful analytical expression for the dependence on the number of fermions, number of components and on interaction strength. Moreover, using a virial approach, we study the Tan's contact behaviour at large temperatures and in the limit of infinite interactions and we show that it increases with the temperature and the number of components. At zero temperature, we predict that the weight of the momentum distribution tails increases with interaction strength and the number of components if the population per component is kept constant. This latter property was experimentally observed in Ref.~[Nat. Phys. {\bf 10}, 198 (2014)].
\end{abstract}


\maketitle


\section{Introduction}
The recent progresses on the experimental control offered by ultra-cold atoms setups~\cite{Bloch2008,Lewenstein2012,Cazalilla2011,Guan2013} provides new platforms for the theoretical and experimental activity on one dimensional (1D) strongly correlated quantum systems. 
In the spirit of quantum simulations~\cite{QuSim2012}, they now allow to study many important phenomena like for instance superfluidity, integrable models~\cite{Guan2013}, quantum phase transitions and quantum magnetism~\cite{Rey2014}.
In the latter case, multi-component strongly interacting quantum particles, living in continuous space, appear to be a promising alternative to lattice systems where magnetic interaction parameters are hardly tunable~\cite{Murmann2015}.
Indeed, in the low-energy regime, the $\kappa$ internal degrees of freedom (interacting via highly-symmetric terms) can be mapped onto an effective $SU(\kappa)$ spin chain subjected to a Sutherland Hamiltonian~\cite{Sutherland68,Takahashi70}, which reduces to the most traditional $SU(2)$ Heisenberg model in the case of two components~\cite{Volosniev2014,Deuretzbacher2014,Decamp2016}.
Such systems are therefore currently at the focus of an intense experimental~\cite{Pagano2014,Taie2010,Murmann2015} and theoretical~\cite{Cui2013,Massignan2015,Grining2015,Beverland2016} activity.
Inevitably, a compelling question arises about probing the magnetic-like properties of these gases, via experimental techniques as elementary as possible, besides \emph{in-situ} spin-resolved imaging of the cloud.
Our purpose is here to endorse the use of the \emph{standard} time-of-flight momentum distribution(s)~\cite{Bloch2008,Lewenstein2012}, and in particular of large-momenta tails thereof, as a complementary diagnostic tool for magnetic-like structures.

The momentum distribution $n(k)$ of a Fermi Gas is indeed a powerful probe of both fermionic statistics and of the intertwined effect of interactions between particles and the effective dimensionality they move in.
For a homogeneous system of non-interacting fermions at zero-temperature, $n(k)$ is a unit step function of the momentum modulus, sharply vanishing at the Fermi surface (or Fermi points in 1D) $|{\bf k}|=k_F$, where $k_F$ is called the Fermi wave-vector.
In dimension larger than one, many-body effects only reduce the jump at the Fermi surface to a value smaller than one~\cite{Huotari2010} and the system falls in the universality class of Fermi-liquids.
In dimension one, instead, the picture is considerably changed and the momentum distribution derivative displays a power-law discontinuity at $k=k_F$, i.e., $n(k) \sim |k-k_F|^\beta$ with an exponent as predicted by the Tomonaga-Luttinger liquid theory~\cite{Haldane81}.
A remarkable common feature in all dimensions, however, is the presence of universal power-law tails $n(k)\sim k^{-4}$ for a gas where interactions can be schematized as contact ones (as it is the case for most standard cold gases experiments). 
The weight of such tails, denoted as Tan's contact, can be put into relation with several many-body quantities, ranging from the interaction energy to the depletion rate by inelastic collisions, and many more~\cite{Tan2008a,Tan2008b,Tan2008c}.
In this work, we show that such tails also encode precious information about the permutational symmetry hiding behind the formation of magnetic-like structures in strongly-interacting multi-component gases.

The presence of a universal Tan's contact is a robust feature related to the interaction form only, but its value is expected to be influenced by the type of confinement the Fermi gas is subjected to, as it is indeed the case for bosons~\cite{Minguzzi02,Olshanii03,vignolo2013}. 
Its value for strongly interacting, homogeneous, two-component Fermi gas has been calculated both in three~\cite{Hui11} and in one dimension~\cite{Zwerger2011}; for a multi-component mixture, the Bethe Ansatz exact solution~\cite{LiebLin,Yang67,Sutherland68,Takahashi70,LaiYang} has been exploited to extract a strong-coupling expansion in the thermodynamic limit~\cite{Guan2012,Patu2016}.
However, despite many cold atomic experimental setups are still based on a harmonic confinement~\cite{Bloch2008,Lewenstein2012}, a few facts are known to date for the correspondent fermionic momentum distribution, besides the smearing of the Fermi sphere due to the inhomogeneous density of the atomic cloud.
Here, we focus on such a situation and we provide theoretical and numerical evidence for a scaling relation between the Tan's contact value, the interaction strength, the harmonic confinement and the number of particles and components (as well as the temperature), thus putting solid grounds for some recent numerical observations~\cite{Matveeva2016}.

In this work, we combine an exact solution at infinite interactions and numerical Matrix-Product-State (MPS) simulations at finite interactions, in order to predict the behavior of the momentum distribution tails for a mixture of $\kappa$ fermionic components under 1D harmonic confinement and interacting among each other with completely $SU(\kappa)$-symmetric repulsive contact interactions.
Moreover, we show that both the ground and excited states of the system have a well defined symmetry thus indicating magnetic-like properties of the mixture~\cite{Volosniev2014,Deuretzbacher2014,Decamp2016,Cui2013,Massignan2015,Grining2015,Deuretzbacher2008} such as a generalized Lieb-Mattis theorem~\cite{LiebMattisPR}:  we claim that such information is fully encoded in the Tan's contact.
Furthermore, we propose a local-density functional approach to determine the Tan's contact for arbitrary particle numbers and number of components  and corroborate the above mentioned scaling relation, which was qualitatively highlighted in the pioneering Florence setup~\cite{Pagano2014} and is amenable of further experimental confirmations also in a few-particle experiments like the ones done in Heidelberg~\cite{Murmann2015}.
Many of the employed techniques and highlighted scalings can then be readily extended to other physical quantities of interest in the wealthy arena of fermionic $SU(\kappa)$ gases.

The paper is organized as follows: Sec. \ref{sec:Model} describes the model and the quantities of interest, momentum distribution and Tan's contact; Sec. \ref{sec:Scaling} puts forward the scaling argumentation to obtain a $N^{5/2}$ dependence of contacts, provides the LDA proof for it and a series expansion at large effective interactions, and displays the numerical evidence for our theoretical findings; Sec. \ref{sec:Magnetic} then presents the asymptotic exact solution and the deep connection between Tan's contacts and internal magnetic structure, a central result of our work, and discusses its robustness to temperature effects; finally, Sec. \ref{sec:Concl} summarizes the main results of our paper and some open perspectives. In addition, several appendixes provide important details about the different methods employed throughout the paper.


\section{Model and quantities of interest}
\label{sec:Model}

\medskip
We consider a $\kappa$ component Fermi gas, made up of $N = \sum_{\nu=1}^\kappa N_\nu$ particles of equal mass $m$, trapped in a tight optical confinement as in the Florence experiment~\cite{Pagano2014}.
The potential is therefore taken as a 1D harmonic trap of frequency $\omega$ and characteristic length $\aho = \sqrt{\hbar / m \omega}$:
\begin{equation} \label{eq:hamfree}
  H_0 = \sum_{i=1}^{N} H^{(\mathrm{1p})}_i \equiv \sum_{\nu=1}^{\kappa}\sum_{j=1}^{N_\nu} \left( - \frac{\hbar^2}{2m} \frac{\partial^2}{\partial x_{j,\nu}^2} + \frac{1}{2} m \omega^2 x_{j,\nu}^2 \right) \ .
\end{equation}

Interactions are dominated by $s$-wave collisions in a  completely $SU(\kappa)$ symmetric channel 
between fermions belonging to different components and therefore the two body interaction potential is accurately described by $v(x-x') = \gd \, \delta(x-x')$, where $\gd = - 2 \hbar^2 / m a_{\mathrm{1D}}$ and $a_{\mathrm{1D}}$ is the 1D effective scattering length~\cite{Olsh98}. The interaction part of the Hamiltonian then reads
\begin{equation} \label{eq:hamint}
  H_{\mathrm{int}} = \gd \sum_{\nu<\nu^\prime}^\kappa H_{\mathrm{int}, (\nu,\nu^\prime)} \equiv
\gd \sum_{\nu<\nu^\prime}^\kappa \sum_{j=1}^{N_\nu} \sum_{j^\prime=1}^{N_{\nu^\prime}} \delta (x_{j,\nu} - x_{j^\prime,\nu^\prime}) .
\end{equation}
The effect of interactions can be recast into a cusp condition 
for each pair of coordinates belonging to different components, $x = x_{j,\nu} - x_{j^\prime,\nu^\prime}$:
\begin{equation}
\label{eq:cusp}
\partial_x \Psi(x=0^+) - \partial_x \Psi(x=0^-) = \frac{2 m \gd}{\hbar^2} \, \Psi(x=0) \ ,
 \end{equation}
with $\Psi = \Psi(X) = \Psi(x_1, \ldots, x_N)$ the many-body wave-function, where we have defined $X=x_1, \dots, x_N=x_{1,1}\dots x_{N_1,1},x_{1,2}\dots x_{N_2,2},
x_{1,\kappa}\dots x_{N_\kappa,\kappa}$.

The momentum distribution is defined as the Fourier transform of the one-body density matrix $\rho_{\nu}(x, x^\prime)$:
\begin{eqnarray}
\label{eq:momentum}
n_{\nu}(k) & = & \frac{1}{2 \pi}\iint \mathrm{d}x\,\mathrm{d}x^\prime\,  \rho_{\nu}(x,x^\prime)e^{-ik(x-x^\prime)} \\
\label{eq:rhonu}
\rho_{\nu}(x, x^\prime) & = & N_\nu\int\mathrm{d}x_2\ldots\mathrm{d}x_N\Psi(X)\Psi(X^\prime) ,
\end{eqnarray}
where $X=(x,x_2,\ldots, x_N)$ and $X^\prime=(x^\prime,x_2,\ldots, x_N)$, with $x, x^\prime$ belonging to the same species $\nu$;
the normalization condition then reads $\int  n_{\nu}(k)\, {\rm d}k=N_\nu$.
In the extreme case of infinite interactions, we  obtain the momentum distribution  from an exact solution for the many-body wave-function.
The latter is based on the mapping of the multicomponent mixture onto a non-interacting Fermi gas with $N$ particles (see Sec. \ref{sec:Magnetic} below and App.~\ref{app:detail-ap}).
In the case of finite interactions, we  search for the ground state of the Hamiltonian  Eqs.~\eqref{eq:hamfree}-\eqref{eq:hamint}  using  a MPS algorithm for an equivalent lattice problem, where we adopted a limiting procedure of the lattice discretization (see App. \ref{app:details-mps}). We then  evaluate the two-point correlators involved in Eq.~\eqref{eq:rhonu} in order to obtain the momentum distribution.
\begin{figure}
  \includegraphics[width=0.9\linewidth]{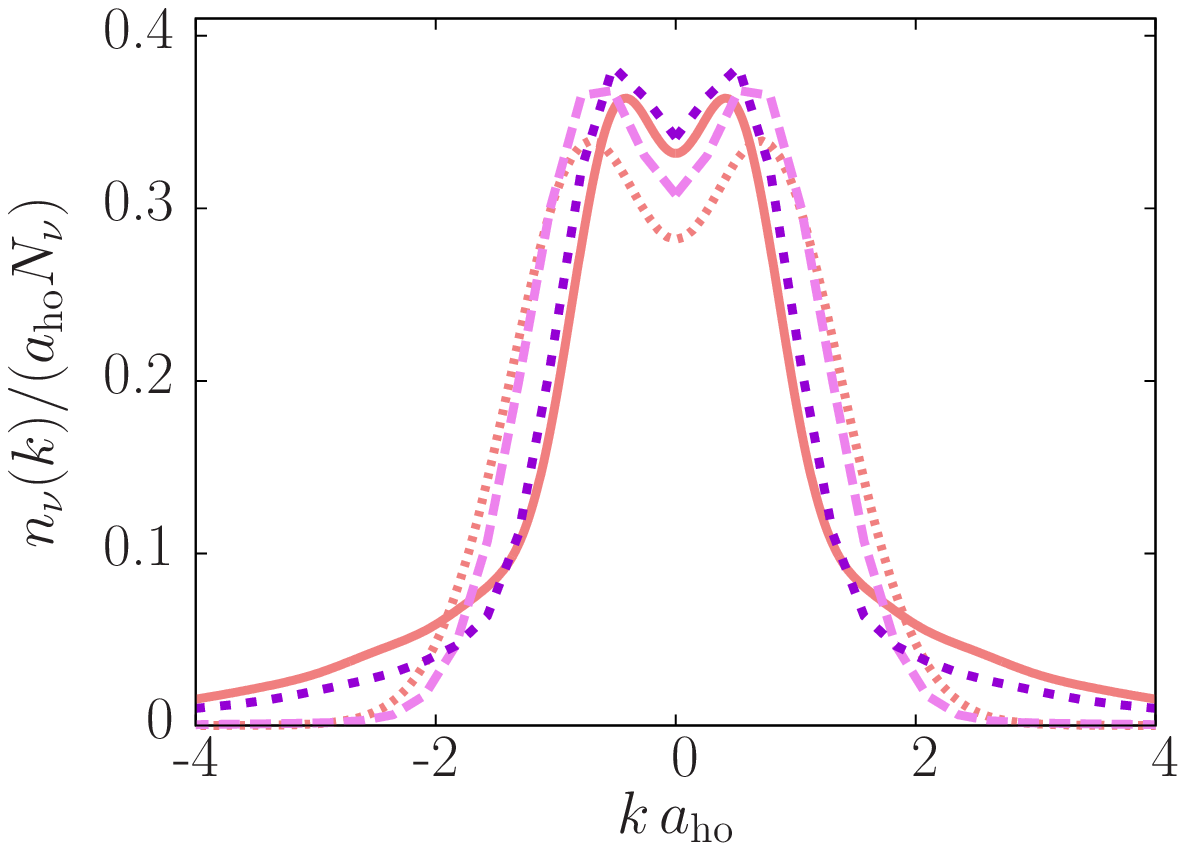}  
  \includegraphics[width=0.9\linewidth]{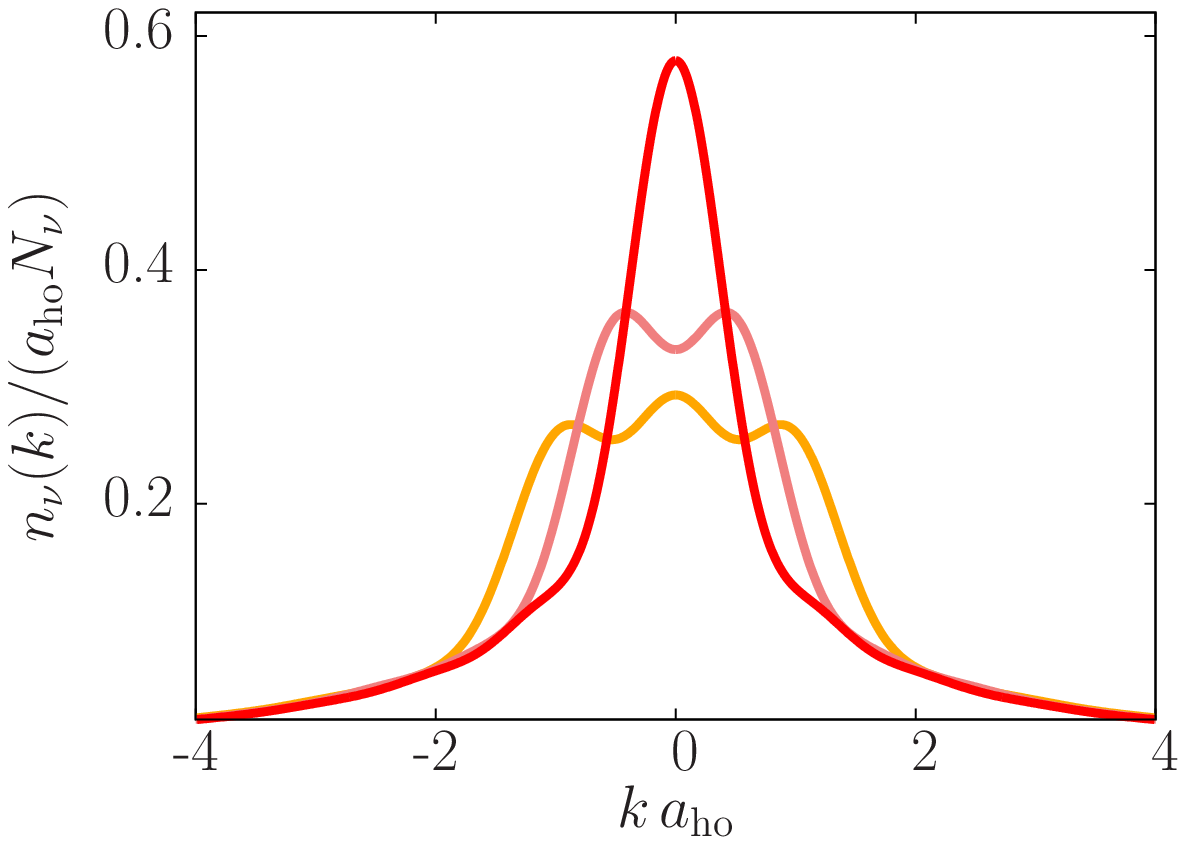}
  \includegraphics[width=0.9\linewidth]{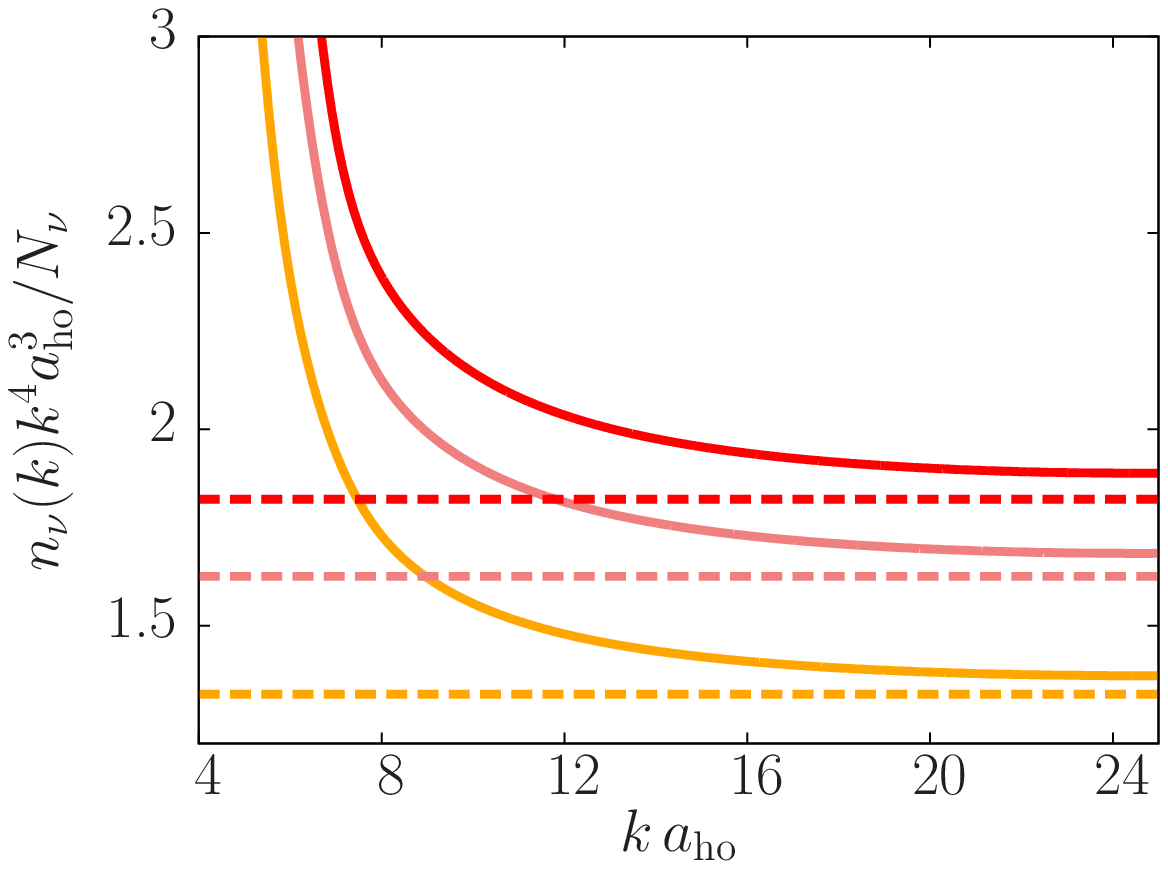}
  \caption{(Color online) Normalized momentum distributions $n_\nu(k)/N_\nu$ (top and middle panels), in units of $\aho$,  and the function $k^4 n_\nu(k)/N_{\nu}$ (bottom panel), in units of $\aho^{-3}$ as a function of $k\aho$, for the case of balanced mixtures with $N=6$. Top panel: all curves correspond to the mixture with $\kappa=3$. The dotted line corresponds to $\gd=0$, the long-dashed line to $\gd=1 \hbar\omega\aho$, the short-dashed line to $\gd=10 \hbar\omega\aho$ and the continuous line to $\gd\to\infty.$ Middle and bottom panels: from the highest curve to the lowest, $\kappa=6$ (red), $\kappa=3$ (coral) and $\kappa=2$ (orange). The horizontal lines in the bottom panel correspond to the exact solution for the contacts given in Eq. (\ref{formulecontact}).\label{fig:mom}}
\end{figure}

The momentum distribution of each species, $n_{\nu}(k)$, is a quantity of primary theoretical and experimental interest,
since it is routinely detected in cold atoms setups via spin-selective time-of-flight techniques, i.e., 
by release of the trapping potential and mapping of the in-trap momenta onto after-flight positions~\cite{Bloch2008,Lewenstein2012}.
The theoretical interest in the momentum distribution resides in the wealth of information that can be extracted therefrom.
In the case of a non-interacting gas, thanks to the duality of spatial and momentum coordinates in the harmonic oscillator Hamiltonian,
$n_{\nu}(k)$ has the same shape as the density profile $n_{\nu}(x) \equiv \rho_{\nu}(x, x)$.
Therefore, it displays a number of peaks coinciding with the number of fermions of that component~\cite{Vignolo00,Mueller2004,Lindgren2014},
with the amplitude of these Friedel-like oscillations decreasing as the inverse of the number of fermions.
Also for an interacting gas, it was found that in a two-species mixture the momentum distribution displays as many peaks as the number of fermions in each component~\cite{Deuretzbacher2016},
although this form does not correspond anymore to the real space distribution.

Among all momentum distribution features, the most striking one  is the power-law behavior at large momenta.
Similarly to the case of a bosonic gas~\cite{Minguzzi02,Olshanii03}, such a $k^{-4}$ behavior of the high-momenta tails of the distribution 
can be directly deduced from the cusp condition~\eqref{eq:cusp}.
The weight of the momentum distribution tails is fixed by the Tan's contact 
$\mathcal{C}_{\nu} = \lim_{k\to_{\infty}} n_{\nu}(k) k^{4}$. 
The Tan's contact is a two-body quantity associated to the interaction energy for a system with contact interaction, that appears in many properties of the many-body system, e.g., in the adiabatic theorem linking the contact to the variation of ground state energy with respect to the {\it s}-wave scattering length, in the virial theorem and in the rate of depletion of the gas by inelastic collisions~\cite{Tan2008a,Tan2008b,Tan2008c,Braaten2012,Zwerger2011}.
In the case of multicomponent mixtures, the Tan's relation reads
\begin{equation}
\label{def:cnu}
\mathcal{C}_{\nu} = \, \frac{ \gd m^2}{2\pi\hbar^4 }\left\langle H_{\mathrm{int},\nu} \right\rangle 
\equiv \, \frac{ \gd m^2}{ 2\pi\hbar^4}\sum_{\nu^\prime \neq \nu}^{\kappa} \left\langle H_{\mathrm{int},(\nu,\nu^\prime)} \right\rangle .
\end{equation}
We notice that, in general, each component has a different Tan's contact,
while for the special cases of two-component mixtures (any $N_1, N_2$)
or of balanced mixture ($p_\nu \equiv N_\nu / N = 1 / \kappa \ \ \forall \nu$)
we get the same contact for all species.
The Tan's contacts are also related to the slope of the energy $K=-\left(\partial E/\partial \gd^{-1}\right)_{\gd \to\infty}$, by virtue of the Hellman-Feynman theorem, as
\begin{equation}
 \mathcal{C}_\mathrm{tot} = \sum_\nu \mathcal{C}_{\nu} =  \frac{m^2}{ \pi\hbar^4}\, K .
\label{eq:gtan2}
\end{equation}

We plot our results for the  momentum distribution $ n_{\nu}(k)$ for a balanced mixture  in Fig.~\ref{fig:mom}.  We first consider the case of a three-component mixture. At increasing interactions, we observe  a narrowing of the momentum distribution in its central part (also noticed for the two-component mixture~\cite{Deuretzbacher2016}),  and an enhancement of the high-momentum tails. As we shall discuss below in Sec. \ref{sec:Scaling}, the increase of the Tan's contact indicates the onset of correlation effects. The middle panel shows the momentum distribution at infinite interactions for various choices of the number of components ($\kappa = 2, 3, 6$, with $N=6$) 
and the lowest panel emphasizes on the momentum distribution tails for the same mixtures by displaying $k^4 n(k)$ . We notice that different mixtures with the same total particle numbers display different values for the Tan's contact.
We will  show below in Sec. \ref{sec:Magnetic} that this can be univocally associated to the different symmetry under permutation, thus disclosing a path for the investigation of magnetic structures via \emph{standard} time-of-flight techniques.


\section{Scaling approach and local-density approximation}
\label{sec:Scaling}

As a first step in our investigations, we analyse the dependence of the Tan's contact of the harmonically trapped mixture on the interaction strength
and total population, deriving a very general scaling approach from a few basic assumptions.
Then we show its validity via a density-functional approach within the local density approximation, still valid for any kind of mixture (balanced or not).
Finally, for the sake of simplicity and closeness to the experiments of Ref.~\cite{Pagano2014}, we provide explicit predictions for the case of a balanced mixture and we compare them with the numerical and analytical results in the trap.

Due to their fermionic nature, the particles  occupy a number of harmonic orbitals which scales with $N$ 
in both extreme regimes of free particles ($\kappa$ independent gases, filling the lowest $p_\nu N$ orbitals each)
and infinitely interacting ones (\emph{as if} there was a single-component gas of $N$ particles filling the lowest $N$ orbitals, as predicted by the exact solution in Sec. \ref{sec:Magnetic} below).
Consequently, we can write the total energy of the system as $E = \hbar \omega \, N^2 \, f(\alpha, N, \{p_\nu\})$, where  $\alpha = 2\aho/|a_{1\mathrm{D}}|$, $p_\nu = N_\nu / N$ and  $f$ a dimensionless function of its arguments (see App.~\ref{app:details-scaling} for details). 
For large systems, the thermodynamic limit  of this expression is achieved by keeping 
the chemical potential at $\alpha \to \infty$, i.e., $\EF = N \hbar \omega$ fixed \cite{Damle1996,Rigol2005}
and therefore rescaling the interaction parameter into $\alpha_0 = \alpha / \sqrt{N}$ (see App. \ref{app:details-scaling} for details).
In this limit we conjecture that 
\begin{equation}\label{eq:enscale}
	E / N = \EF f(\alpha_0, \{p_\nu\}) = N \, \hbar \omega \, f(\alpha / \sqrt{N}, \{p_\nu\})
\end{equation}
From the scaling expression of Eq.~(\ref{eq:enscale}) we can easily derive the expected scaling for the Tan's contact,
by using its definition (\ref{eq:gtan2}) and the relation $\gd = \hbar \omega \aho \alpha_0 \sqrt{N}$:
\begin{equation}\label{eq:Cscale}
	{\mathcal C}_\mathrm{tot}(\alpha_0) 	= \frac{1}{\pi \aho^3} N^{5/2} \, \alpha_0^2 \frac{\partial f(\alpha_0, \{p_\nu\})}{\partial \alpha_0} ,
\end{equation}
and similar ones for the single $\mathcal{C}_\nu$'s.
The above equation shows that the contact displays a scaling behavior as a function of the parameter $\alpha_0$.  Such a behavior has been recently noticed in numerical Monte Carlo simulations by~\cite{Matveeva2016}:
we provide here a theoretical ground for this observation.

We show in fact (details in App.~\ref{app:details-balda}) that Eq.~(\ref{eq:enscale}) holds exactly within a density functional approach
under a local density approximation (LDA) based on the Bethe-Ansatz (BA) solution of the homogeneous problem~\cite{Guan2012},
$E/ L = (\hbar^2 / 2 m) \rho^3 e(\gamma)$, with $\gamma = m \gd / \hbar^2 \rho$.
This yields indeed an equation for the inhomogeneous density profile $\rho(x)$ in a external potential $V_\mathrm{ext}(x) = m \omega^2 x^2 / 2$,
\begin{equation}
\frac{3}{2} \frac{\hbar^2}{m} \rho^2 e(\gamma) - \frac{1}{2} \gd \rho e'(\gamma) = \mu - V_\mathrm{ext}(x)
\label{rholda-main}
\end{equation}
as well as an equation for the contact
\begin{equation}
{\mathcal C}_\mathrm{tot} = \frac{\gd^2}{2\pi} \int dx \rho^2(x) e'(\gamma),
\label{clda-main}
\end{equation}
which properly rescaled by the typical length $\aho$ and energy $\hbar \omega$ lead to the above Eqs.~\eqref{eq:enscale}-\eqref{eq:Cscale}.
In particular, in the regime of strong interactions we resort to a strong coupling series' expansion for the homogeneous Bethe-Ansatz energy functional~\cite{Guan2012} (see App. \ref{app:details-balda}).
For a balanced mixture we obtain the LDA expression for the Tan's contact at infinite interactions
\begin{equation}
\label{eq:CbaldaINF}
{\mathcal C}_\nu(\infty)= \frac{1}{\aho^3} \frac{128 \sqrt{2}}{45 \pi^3} \frac{Z_1(\kappa)}{\kappa} N^{5/2} ,
\end{equation}
as well as at finite large interactions, up to $O(1/\alpha_0^2)$:
\begin{widetext}
\begin{equation}
\label{eq:Cbaldafiniteg}
\mathcal{C}_\nu (\alpha_0)  =  
\frac{N^{5/2}}{\pi \kappa \aho^3}
\left[ 
	\frac{128 \sqrt{2} Z_1(\kappa)}{45 \pi^2}
	+ \frac{2 (315 \pi^2 - 4096) Z_1(\kappa)^2}{81 \pi^4 \alpha_0}
	- \frac{64 \sqrt{2} \left(25 (1437 \pi^2 - 14336) Z_1(\kappa)^3 + 1728 \pi^4 Z_3(\kappa)\right)}{14175 \pi^6 \alpha_0^2}
\right]
\end{equation}
 
\end{widetext}
where $Z_1(\kappa) = -\frac{1}{\kappa}(\psi(\frac{1}{\kappa}) + C_\mathrm{Euler})$, $Z_3(\kappa)=[\zeta(3,1/\kappa)-\zeta(3)]/\kappa^3$, $\psi$ being the digamma function, $C_\mathrm{Euler}$ the Euler constant and $\zeta$ the Riemann functions. We stress here that in the limit of infinite numbers of components~\cite{YangYou2011}, recalling that $\lim_{\kappa \to \infty} Z_1(\kappa) =1$, we readily  obtain the expression for the bosonic Tan's contact in LDA generalizing the result of Ref. \cite{Olshanii03} for $\alpha_0\to\infty$.
Moreover, this expression also generalizes the results of Ref. \cite{Grining2015} for a two component Fermi gas (Gaudin-Yang model in a harmonic trap) and complement the ones of Ref. \cite{Zinner2016} for the polaron case.  

\begin{figure}
  \includegraphics[width=1\linewidth]{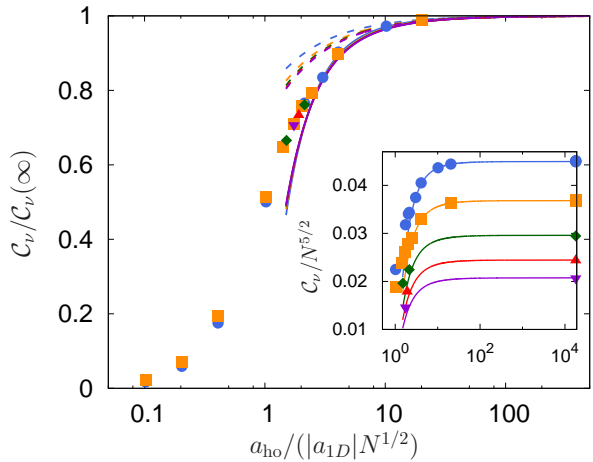}
  \caption{\label{fig:datacollapse}(Color online) Exact and MPS results for $\mathcal{C}_\nu$ for balanced mixtures with different number of fermions (up to $N=12$), components, and interaction strengths, as functions of the parameter $\alpha_0/2=\aho/(|a_{1\rm{D}}|N^{1/2})$. The blue circles correspond to $\kappa=2$, the orange squares to $\kappa=3$, the green diamonds to $\kappa=4$, the red up-triangles to $\kappa=5$, and the violet down-triangles to $\kappa=6$. The data are compared with the perturbative LDA expression~(\ref{eq:Cbaldafiniteg}) to first- and second-order in $1/\alpha_0$ (dashed and continuous lines, respectively). The inset shows the data collapse on the same $\kappa$-dependent curve when the $N^{5/2}$ dependence of $\mathcal{C}_\nu$ is taken away; the main panel then displays the very weak dependence on $\kappa$ once the limiting value at $\gd \to \infty$ of Eq.~\ref{eq:CbaldaINF} is also factorized away.
}
\end{figure}

Fig.~\ref{fig:datacollapse} shows the results for the Tan's contact at various interaction strengths as well as for various numbers of components as obtained by  the exact solution in the infinitely repulsive limit (see Sec. \ref{sec:Magnetic} and App. \ref{app:detail-ap}), the numerical MPS approach (see App. \ref{app:details-mps} for details) and the LDA. Quite remarkably, the LDA is in excellent agreement with the numerically exact data at strong interactions, where the Bethe-Ansatz series expansion holds. In order to properly describe weaker interactions, one would need higher-order terms in the Bethe-Ansatz equation of state. 
We also find that  the scaling of Eq.~(\ref{eq:Cscale}) is satisfied to a high degree of precision by the numerical MPS simulations in all interaction regimes, and the dependence on the number of components is essentially captured for homogeneous mixtures by the prefactor $Z_1(\kappa) / \kappa$ in Eq.~(\ref{eq:CbaldaINF}).

By virtue of Eq.~(\ref{eq:CbaldaINF}) and the numerical observation in Fig.~\ref{fig:datacollapse}, we then provide an analytical prediction (even at finite interactions) for the growth of the $n_\nu(k)$ tails with the number of components $\kappa$, by keeping fixed the population per component $N_\nu$:
\begin{equation}
  \label{eq:CbaldaLENS}
  C_\nu \propto N_\nu^{5/2} Z_1(\kappa) \kappa^{3/2}.
\end{equation}
This prediction qualitatively explains the experimental observation~\cite{Pagano2014} of an increase of the momentum distribution tails with increasing number of components \cite{noteexp}. A direct comparison to experimental data would be made difficult by the (yet) unavoidable convolution average among several 1D tubes with different total population: however, provided the balanced nature of the mixture is preserved across the sample, the dependence on the number of components $\kappa$ factorizes out of the integrals and would be readily verifiable.
%


\section{Symmetry spectroscopy}
\label{sec:Magnetic}

{\it Exact solution and ground state degeneracy.---}
In the limit $\gd\to\infty$  the gas further \emph{fermionizes} due to repulsion~\cite{Zurn2012,Volosniev2013,Volosniev2014}, and the ground and first excited states merge into a vastly degenerate manifold. This degeneration is related to the \emph{emergent} $S_N$ symmetry under permutation of all the particles among each other~\cite{Harshman2014,Harshman2016a,Harshman2016b}. To obtain the many-body wave-functions for the manifold we start from the totally antisymmetric wave-function $\Psi_A(x_1,\dots,x_N)=\frac{1}{\sqrt{N!}}\mathrm{det}[\phi_{i-1}(x_j)]_{i,j=1,\ldots,N}$, which satisfies by construction the cusp condition (\ref{eq:cusp}).  Here $\phi_0,\ldots,\phi_{N-1}$ are the eigenfunctions of the single particle Hamiltonian $H^{(\mathrm{1p})}_j$. We then write the full many-body wave-function as~\cite{Volosniev2014,Deuretzbacher2014,Decamp2016}:
\begin{equation}
\label{eq:Psi}
\Psi(X)=\sum_{P\in S_N}a_P\chi_P(X)\Psi_A(X),
\end{equation}
where $X=(x_1,\ldots,x_N)$, $S_N$ is the permutation group of $N$ elements and $\chi_P(X)$ is equal to $1$ if $x_{P(1)}<\cdots<x_{P(N)}$ and $0$ otherwise. Imposing the antisymmetry condition to the fermions belonging to each given component, we have that the number of independent $a_P$ coefficients  is given by $D_{N,\kappa}=N!/(N_1!...N_\kappa!)$, which corresponds also to  the dimension of the degenerate ground-state manifold (often called a ``snippet'' basis). In order to identify within the manifold the wave-function which corresponds to the unique ground state at finite, large interactions, we use  degenerate perturbation theory to order $1/\gd$, and determine the set of coefficients $a_P$  which maximizes $K=-\left(\partial E/\partial \gd^{-1}\right)_{\gd \to\infty}$~\cite{Volosniev2014,Decamp2016} (see App. \ref{app:detail-ap} for details).

{\em Symmetry identification.---}
The underlying symmetry is the equivalent of magnetic structures for spinor systems in the sense that the effective interaction Hamiltonian (restricted to the first degenerate manifold (see App. \ref{app:detail-ap})) can always be mapped onto a spin chain Hamiltonian \cite{Deuretzbacher2014,Volosniev2014,Sutherland68,Takahashi70}. The permutation symmetry is at the heart of the fermionized solution (\ref{eq:Psi}): all the particles may be permuted and occupy all the single particle orbitals.
Here we extend the results of~\cite{Decamp2016} and show that each state of the  many-body wave-function in Eq.(\ref{eq:Psi}) can be associated to a single Young Tableau, or equivalently a single symmetry,
and that the ground state corresponds to the most symmetric configuration compatible with the imbalance.
This constitutes a generalization of the well known Lieb-Mattis's theorems 
about magnetic ordering of 1D two-component fermions and spin
chains~\cite{LiebMattisPR}.

Furthermore, considering various possible mixtures at fixed total number of fermions, the value of the interaction energy parameter $K$ is the largest for the most symmetric mixture, i.e. the one with the largest number of components. Using these facts together with Eq.(\ref{eq:gtan2}), we  obtain that the most symmetric mixture  will display the largest tails in the momentum distribution. This is confirmed by the exact calculation of the momentum distribution, as illustrated in Fig.~\ref{fig:mom}. 

\begin{figure}
  \includegraphics[width=0.9\linewidth]{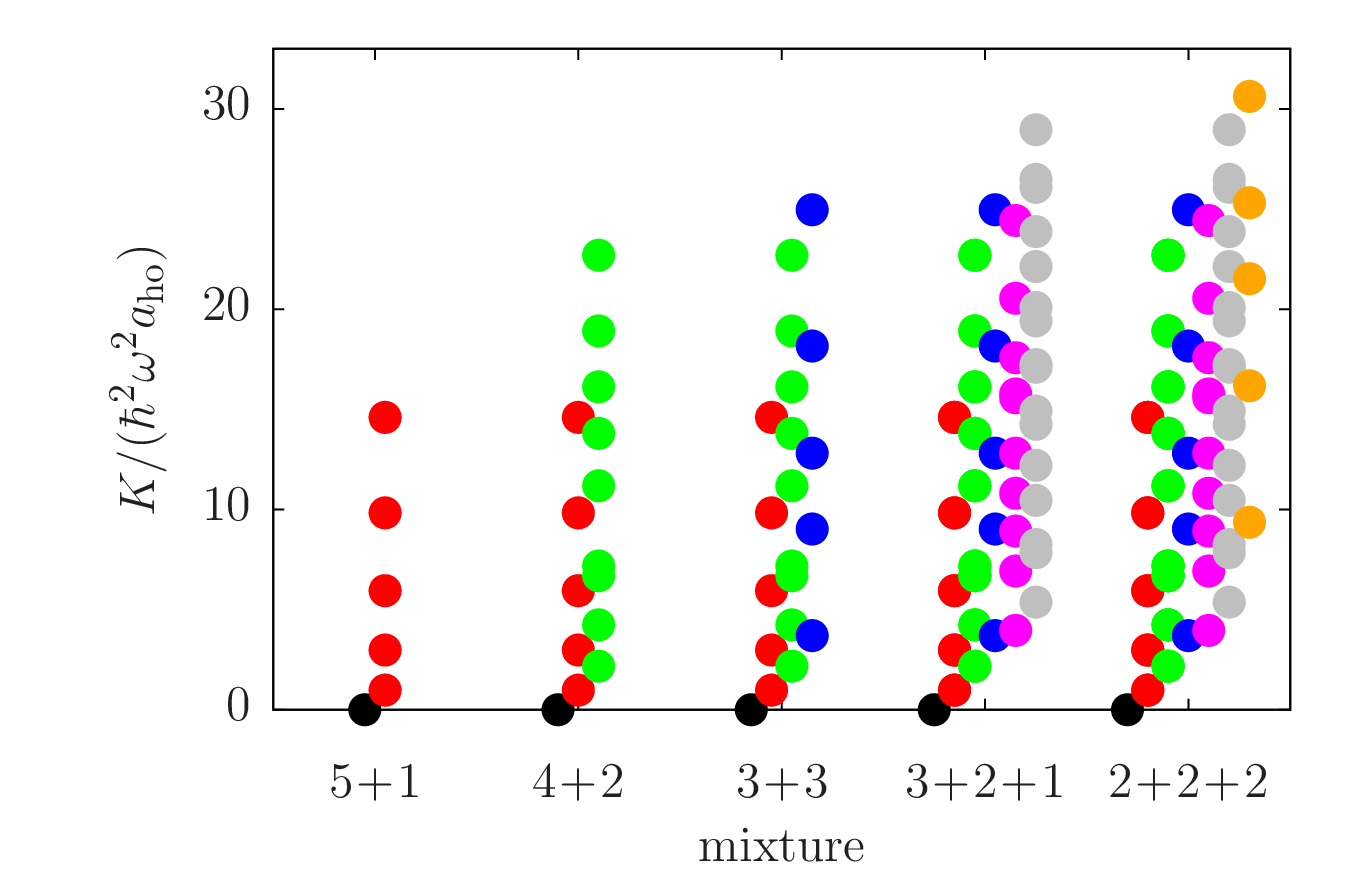}
  \caption{\label{fig:sym}(Color online) The interaction energy parameter $K$ (or equivalently $\mathcal C_{\rm tot}$) for different mixtures: 5+1, 4+2, 3+3, 3+2+1, and 2+2+2. The different colors correspond to different symmetries characterized by their Young's tableaux defined in the main text and Table~\ref{tableau_y}: $Y_{-15}$ (black), $Y_{-9}$ (red), $Y_{-5}$ (green), $Y_{-3,-8}$ (blue), $Y_{-3,4}$ (magenta), $Y_{0}$ (gray), $Y_{3}$ (orange). }
\end{figure}

The relation between the $K$ parameter or equivalently the contact $\mathcal C_{\rm tot}$ (see Eq.~(\ref{eq:gtan2})) and the symmetry of the corresponding state is shown in Fig.~\ref{fig:sym} for the case of different mixtures. For a specific mixture, the symmetry of each state is labeled on the one hand by calculating the expectation value of the transposition class-sum operator $\Gamma^{(2)}=[(2)]_N=\sum_{i<j}(ij)$~\cite{Katriel1993,Novolesky1994,Fang2011}, and on the other hand by constructing all the Young tableaux compatible with the Pauli principle in the mixture (see App. \ref{const-young}). Indeed, each Young tableau $Y_{\gamma_2}$ corresponds to a unique eigenvalue $\gamma_2$ of $\Gamma^{(2)}$, via the expression
\begin{equation}
  \gamma_2=\dfrac{1}{2}\sum_i\lambda_i(\lambda_i-2i+1),
  \label{young}
\end{equation}
where $i$ and $\lambda_i$ refer respectively to the line and number of boxes in this line of Young tableau. For the cases when $\gamma_2$ corresponds to different Young tableaux, we calculate the expectation value of the $3$-cycle class sum operator $\Gamma^{(3)}$~\cite{Katriel1993} too, and we univocally identify the corresponding Young tableaux $Y_{\gamma_2,\gamma_3}$ by means of the eigenvalues $\gamma_2$ and $\gamma_3$ related to the operators $\Gamma^{(2)}$ and $\Gamma^{(3)}$. Note that the degeneracies of the couples ${\gamma_2,\gamma_3}$ as eigenvalues of the class-sum operators are equal to the number of states with the corresponding symmetry, and are also equal, in terms of group theory, to the dimension of the corresponding irreducible representation \cite{Hamermesh_book}.
 
\begin{table}
  \begin{center}
    \begin{tabular}{p{0.3\linewidth}p{0.3\linewidth}p{0.3\linewidth}}
      \hline \hline \\
      \textcolor{black}{$Y_{-15}=\;\scriptstyle{\yng(1,1,1,1,1,1)}$} & \textcolor{red}{$Y_{-9}=\;\scriptstyle{\yng(2,1,1,1,1)}$} & \textcolor{green}{$Y_{-5}=\;\scriptstyle{\yng(2,2,1,1)}$}\\[0.3\linewidth]
      \\
      \textcolor{blue}{$Y_{-3,-8}=\;\scriptstyle{\yng(2,2,2)}$}& \textcolor{magenta}{$Y_{-3,4}=\;\scriptstyle{\yng(3,1,1,1)}$} & \\[0.3\linewidth]
      \\
      \textcolor{gray}{$Y_{0}=\;\scriptstyle{\yng(3,2,1)}$} & \textcolor{orange}{$Y_{3}=\;\scriptstyle{\yng(3,3)}$}&\\[0.2\linewidth]
      \hline \hline
    \end{tabular}
  \end{center}
  \caption{\label{tableau_y}(Color online) Young's tableaux $Y_{\gamma_2,\gamma_3}$ corresponding to the eigenvalues $\gamma_2$ and $\gamma_3$. Boxes belonging to the same line (resp. column) correspond to a symmetric (resp. antisymmetric) exchange of particles. The color code is the same than in Fig. \ref{fig:sym}.}
\end{table}

The graphics in Fig.~\ref{fig:sym} show that the (interaction) energy slope $K$, and thus the total contact $\mathcal{C}_\mathrm{tot}$, are determined by the state symmetry, regardless of the number of components and fermions per component, at fixed number of particles. Indeed, the sets of $K$-values corresponding to one specific Young Tableau are the same for each mixtures, as long as this symmetry is possible. An interesting open problem would be to quantitatively link the values of $(\gamma_2,\gamma_3)$ to the energy slope $K$, further elucidating the information content of the symmetries. 
Independently of the analytic relation, however, we can already highlight the deep relation between the Tan's contact(s) and the magnetic structure of the multi-component $SU(\kappa)$ fermionic gas: although the momentum distribution form does not anymore reflect the spatial density (as commented before), the tails encode some even more precious information about the spin arrangement.

At finite interactions, the symmetry under permutation is only approximately satisfied since the cusps among fermions belonging to different components are not equivalent to the nodes in the fermionic many-body wave-function for particles belonging to the same component. Nevertheless, the good agreement of the numerical results  as compared with the ones  at infinite interactions, both for the momentum distribution function (see Fig.~\ref{fig:momentum_tail}), as well as of the density profiles at interaction strength $\alpha_0\ge 40$, (see App. \ref{details-nx}) suggests that the consequences of the permutation symmetry persist at finite, large interactions. We therefore put forward the precise measurement of momentum distribution tails via standard time-of-flight as a novel diagnostic tool (complementary to \emph{in-situ} cloud imaging) in the yet largely unexplored arena of quantum magnetism.

{\em Temperature dependence on the Tan's contact.---}
At finite temperature, for large interaction strength $\gd$, one may identify two different temperature regimes: at temperature lower than the energy level splitting of the quasi-degenerate ground-state manifold, scaling as $\Delta E \sim K/\gd$, one may recover a behavior close to the zero-temperature one, where only the lowest-energy state is considerably populated, and the mixture has a well defined symmetry. This regime corresponds to the spin-coherent regime described for a two component Fermi gas~\cite{Cheianov2005}.
For temperatures higher than $\Delta E$ the whole manifold is thermally populated, and  the state of the system is described as an incoherent mixture with various symmetry components. For this high-temperature regime we estimate the Tan's contact for the gas using a thermodynamic form of the Tan's relation~\cite{Tan2008c,Zwerger2011,Hui11}, 
\begin{equation}
  \label{eq:tanT}
  \left(\frac{\mathrm d\Delta\Omega_{\nu}}{\mathrm da_{1\mathrm{D}}}\right)_{\mu,T}=\frac{\pi \hbar^2}{m}\mathcal{C}_{\nu},
\end{equation}
where  $\Delta \Omega_\nu$ is the contribution of the $\nu$ component of the mixture to the grand thermodynamic potential $\Omega=\Omega^{(1)}+ \sum_\nu \Delta \Omega_\nu$. Using a virial expansion for $\Delta \Omega_\nu$ we obtain the high-temperature behavior for the contact of a multicomponent fermionic mixture in the limit $\gd \to \infty$ (see App. \ref{app:detail-ap} for details)
\begin{align}
  \begin{split}
    \label{eq:cnuhight}
    \mathcal{C_{\nu}}&=\frac{1}{\pi^{3/2} \aho^3} \sqrt{\frac{\kB T}{\hbar\omega}} N^{2} p_\nu (1-p_\nu)\\
    &=\frac{1}{\pi^{3/2} \aho^3} \sqrt{\frac{\kB T}{\EF}} N^{5/2} p_\nu (1-p_\nu)
  \end{split}
\end{align}
It is important to notice that, as in the bosonic case ~\cite{vignolo2013}, in the fermionized limit the mixture displays universal properties, analogue to those of a unitary Fermi gas in three dimensions. This is seen in the fact that the  second virial contact coefficient $c_2$  (see App. \ref{app:detail-ap} for details) is a constant: $c_2=1/\sqrt{2}$.

\begin{figure}
  \includegraphics[width=0.9\linewidth]{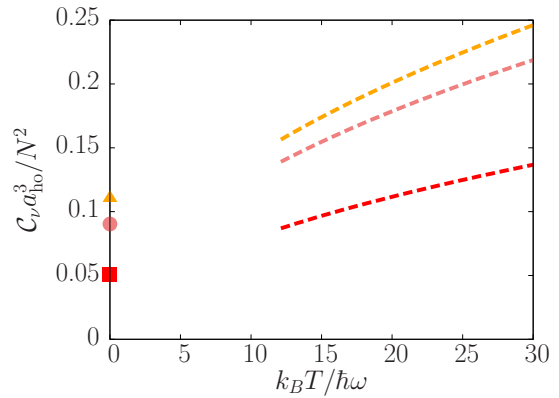}
  \caption{(Color online) Contact $\mathcal{C}_{\nu}/N^2$, in units of $\aho^{-3}$, as a function of the temperature $T$, in units of $\hbar\omega/\kB $, for the case of balanced mixtures with $N=6$. The lines correspond to large temperature behavior given Eq.~\eqref{eq:cnuhight} and the points to the exact solution at $T=0$ given in Eq.~\eqref{eq:CbaldaINF}. From top to bottom: $\kappa=6$ (red), $\kappa=3$ (coral) and $\kappa=2$ (orange).
  }
\label{fig:finiteT}
\end{figure}

The temperature dependence of the Tan's contact for a multicomponent mixture in harmonic trap is shown in Fig.~\ref{fig:finiteT}. In this high temperature regime, for all mixtures, the contact increases with temperature: this is a phase-space effect due to the fact that the mixtures are one-dimensional (see App. \ref{app:detail-ap} for details). Secondly, we remark that also at finite temperature the contact increases with the number of components of the mixture, again in agreement with the experimental observations~\cite{Pagano2014}. In the high-temperature regime, balanced mixture, from Eq.~\eqref{eq:cnuhight} one has $\mathcal{C}_\nu\propto N_\nu^2 (\kappa-1)$. This  dependence of the contact on the number of components is different from the quantum-degenerate case given by Eq.~\eqref{eq:CbaldaLENS}, further  emphasizing the relation between  the contact and the interaction energy. As a last remark, let us underline that if the temperature is rescaled by $\EF$, $\mathcal{C}_\nu$ scales as $N^{5/2}$ as in the zero-temperature regime [Eq. (\ref{eq:cnuhight})].

\section{Conclusions}
\label{sec:Concl}

In this work we have studied the high-momentum tails of the momentum distribution, an observable of primary experimental interest, for the case of strongly correlated multicomponent fermions under harmonic confinement. The weight of the tails, denoted as the Tan's contacts, conveys a wealth of information on the many-body system under study. In particular, the Tan's contacts are fingerprints of fermionization effects as the weight of the tails increases rapidly as the strongly correlated regime is entered.
Using a local-density approach in the Bethe-Ansatz equation of state, which agrees surprisingly well with the numerical MPS results, we have demonstrated and checked on the numerical results that the Tan's contacts scales as a function of the parameter $\alpha_0= 2 a_{ho}/|a_{1\mathrm{D}}|\sqrt{N}$. This scaling allows us also to predict the behavior of the momentum distribution tails as a function of the  number of components of the gas at fixed population, both at zero and at finite (large) temperature. In both regimes we predict an increase of the tails with the number of components which is observed in the Florence experiment~\cite{Pagano2014}. Our analytical approach also provides a theoretical explanation for the $N^{5/2}$ scaling of the Tan's contact observed in Monte Carlo simulations~\cite{Matveeva2016}.

Multicomponent Fermi gases at strong repulsive interactions display an enhanced permutation symmetry stemming from the further fermionization induced by interactions. The ground and excited states of the system in particular can be characterized by unique Young Tableaux, indicating magnetic-like properties of the mixture and satisfying a generalized Lieb-Mattis theorem~\cite{LiebMattisPR,Decamp2016}. We show in this work that, at zero temperature and in the limit of infinite interactions, there is a one-to-one correspondence between the weight of the tails and the symmetry of the mixture. This allows a ``symmetry spectroscopy'' approach through an accurate measurement of the momentum distribution tails, thereby making  abstract concepts of group symmetries directly detectable in experiments.
Even if the precision required to verify our predictions in terms of Young Tableaux might be out of reach to date for large populations, a stringent benchmark could be readily offered by experiments building quantum magnetic simulators in a ``bottom-up'' approach like the ones of Ref.~\cite{Murmann2015}.
On the other hand, the a priori knowledge of the symmetry of the ground state wave-function could considerably lighten the complexity of the numerical simulations~\cite{Mila2014}.

Our results therefore pave the way to a wide range of further theoretical and experimental studies about the profound interplay between permutation group symmetries and strong-correlation effects, at the heart of the microscopic origin of magnetic-like structures. We believe that this will constitute an ideal testbed for quantum simulations with ultracold gases.

\section*{Acknowledgments} 

We acknowledge discussions with T. Busch, L. Fallani, A. Foerster and M. Gattobigio and useful comments from P. van Dongen. AM acknowledges financial support from ANR projects Mathostaq (ANR-13-JS01-0005-01) and SuperRing (ANR-15-CE30-0012-02). During parts of this work, JJ was a recipient of a DFG-funded position through the Excellence Initiative by the Graduate School Materials Science in Mainz (MAINZ - GSC 266). JJ also thanks Studienstiftung des Deutschen Volkes for financial support. The MPS simulations were run by JJ and MR on the Mogon cluster of the JGU (made available by the CSM and AHRP), with a code based on a flexible Abelian Symmetric Tensor Networks Library, developed in collaboration with the group of S. Montangero at the University of Ulm.
%


\appendix


\section{Details on the exact solution}
\label{app:detail-ap}

\textit{Determination of the ground-state wave-function.---}
In order to find the ground-state at infinitesimal $1/\gd$, we employ a degenerate perturbation theory at first order,
starting from the wave-function ansatz of Eq.~\eqref{eq:Psi}, where both $\Psi_A$ and $\Psi$ are normalized to unity.
Using the definition  $K = - \partial E/\partial \gd^{-1}$ and the Hellmann-Feynman theorem we get
\begin{equation}
\label{eq:Kpert}
K=\sum_{P,Q \in S_N} (a_P - a_Q)^2 \alpha_{P,Q} \ ,
\end{equation}
where the matrix element is the integral
$\alpha_{P,Q} = \alpha_k = \int \mathrm{d}x_1\ldots\mathrm{d}x_N \chi_{1\ldots N} \delta(x_k~-~x_{k+1}) \left[\partial\Psi_A/\partial x_k\right]^2$
if $P$ and $Q$ are equal up to a transposition $\tau_k$ of two consecutive coordinates at positions $k$ and $k+1$ belonging to different spin components,
and zero otherwise.
Note that $K = \gd\left\langle H_\mathrm{int} \right\rangle$ where $H_\mathrm{int}=\gd \sum_{i<j} \delta(x_i-x_j)$ is the interaction part of the Hamiltonian,
so that $\alpha_k$ can be seen as the contribution of an exchange at positions $k$ and $k+1$ to the interaction energy.
Once these weights are computed (see also~\cite{Deuretzbacher2014}), 
we obtain the ground state energy and wave-function by maximizing $K$ with respect to the $a_P$ coefficients,
i.e., finding the minimal eigenvalue of the matrix $\alpha_{P,Q}$.
We then use the correspondent eigenvector to study the equilibrium properties of the mixture and in particular its  momentum distribution.
We also notice that this methods allows also to obtain all the excited states corresponding to the degenerate manifold at $\gd=\infty$ by considering 
the higher eigenvalues of the matrix $\alpha_{P,Q}$~\cite{Decamp2016} 
and getting the other $K$ values plotted in Fig.~\ref{fig:sym}.


\textit{Determination of the one-body density matrix.---}
It is convenient to organize the permutations in the following way:
each $P \in S_N$ is denoted $P_{i_k}$ where $i$ is the position of the first particle
and $k\in \{ 1,\ldots,(N-1)! \}$ denotes the permutation of the remaining $N-1$ particles.
We can write $\Psi$ as $\Psi(X)=\sum_{i=1}^N\sum_{k=1}^{(N-1)!}a_{i_k}\chi_{P_{i_k}}(X)\Psi_A(X)$.
Let us suppose that $x_1\le x_1'$, which we can do since the one-body density matrix is symmetric, $\rho(x_1,x_1')=\rho(x_1',x_1)$.
We observe that
\begin{equation}
\rho_{\nu}(x_1,x_1')=N_\nu \sum_{1\le i\le j\le N}\rho_{\nu}^{(ij)}(x_1,x_1') \ ,
\end{equation}
where $\rho_{\nu}^{(ij)}(x_1,x_1')$ is the integral in \eqref{eq:rhonu},
choosing the limits of the integral such that $(i-1)$ coordinates (resp. $(j-1)$) of $\{x_2,\ldots,x_N\}$ are smaller than $x_1$ (resp. $x_1'$).
Moreover, noticing than $\int\mathrm{d}x_2\ldots\mathrm{d}x_N\chi_{P_{i_k}}(X)\chi_{P_{i_l}}(X')\cdots\ne 0$ only if $k=l$,
using the permutational symmetry of $\Psi_A$ and its expression as a Vandermonde determinant~\cite{Forrester03,Decamp2016}, we finally get
{\small
\begin{eqnarray}
\rho_{\nu}(x_1,x_1') & = & N_\nu G_N\sum_{1\le i\le j\le N}C_{ij}\sum_{P,Q\in S_{N-1}}\epsilon(P)\epsilon(Q)\\
\nonumber & \displaystyle{\prod_{l=2}^{N} } & \int_{L_{ij}}^{U_{ij}} \mathrm{d}z    (z-x_1)(z-x_1') \phi_{P(l)-1}(z) \phi_{Q(l)-1}(z) \ ,
\end{eqnarray}
}%
where $G_N=\frac{2^{N-1}}{\sqrt{\pi}N!(N-1)!}$, $C_{ij}=\frac{\sum_{k=1}^{(N-1)!}a_{i_k}a_{j_k}}{(i-1)!(j-i)!(N-j)!}$
and $(L_{ij},U_{ij})=(-\infty,x_1)$ if $l\le i$, $(x_1',+\infty)$ if $l>j$ and $(x_1,x_1')$ otherwise.


\textit{Determination of the Tan's contacts.---}
Similarly to Eq. \eqref{eq:Kpert}, the contact for each species can also be expressed in terms of the coefficients introduced in Eq. \eqref{eq:Kpert}, by
\begin{equation}
\label{formulecontact}
\mathcal{C}_{\nu}=\frac{1}{2\pi}\sum_{\substack{\mu=1 \\ \mu\ne\nu}}^\kappa \sum_{k=1}^{N-1}\sum_{P\in \sigma_N(\mu,\nu,k)}(a_P-a_{(\tau_k\circ P)})^2\alpha_k,
\end{equation}
where $\sigma_N(\mu,\nu,k)$ is the set of all permutations such that particles in $k$ and $k+1$ positions are from the species $\mu$ and $\nu$, and $\tau_k$ is the transposition of these two particles.
We remark here that Eq. (\ref{formulecontact}) naturally generalizes the one obtained in Ref.~\cite{vignolo2013} for a Tonks-Girardeau gas.


\textit{Determination of the large-temperature behavior for the contact.---}
 In order to obtain the finite temperature behavior of the contacts $\mathcal{C}_{\nu}$ via Eq.~\eqref{eq:tanT}, we evaluate  $\Delta \Omega_\nu$  by a virial approach~\cite{vignolo2013,Hui11}, by writing the grand thermodynamic potential as a function of the grand partition function $\mathcal{Z}$, $\Omega=-\kB T\ln\mathcal{Z}$, $\kB $ being the Boltzmann constant and $T$ the temperature. The first step is to make a second order expansion of $\mathcal{Z}$ in terms of the fugacities $z_{\nu}= \exp (\mu_\nu N_\nu / \kB T)$,  which tend to zero at high temperature ($\mu_\nu$  being the chemical potential of the $\nu$ species),
\begin{equation}
  \mathcal{Z}=1+Q_{1,0}\sum_{\nu}z_{\nu}+\left(Q_{2,0}\sum_{\nu}z_{\nu}^2+Q_{2,1}\sum_{\mu\neq\nu}z_{\mu}z_{\nu}\right) \ ,
\end{equation}
where $Q_{n,n^\prime}$ is the partition function of a cluster containing $n-n^\prime$ fermions of species $\nu$
and $n^\prime$ of species $\mu$~\cite{Liu2010}.
Thus the interacting components of the grand thermodynamic potential can be written
\begin{equation}
\Delta\Omega_{\nu} = -2\kB T  Q_{2,1}z_{\nu}\sum_{\mu\ne\nu}z_{\mu}
\end{equation}
and, by using Eq. \eqref{eq:tanT}, the contact for species $\nu$ takes the form
\begin{equation}
\label{ct1}
\mathcal{C_{\nu}}=\frac{4Q_{1,0}}{\LdB^3}c_2z_{\nu}\sum_{\mu\ne\nu}z_{\mu},
\end{equation}
where $\LdB=\sqrt{2\pi\hbar^2/m\kB T}$ is the thermal de Broglie wavelength and
\begin{equation}
  c_2=-\frac{\partial (Q_{2,1}/Q_{1,0})}{\partial (a_{1\mathrm{D}}/\LdB)}
\end{equation}
is the dimensionless second virial coefficient~\cite{Hui11}.
As we have already shown in Ref.~\cite{vignolo2013}, in the strongly interacting regime and in the high-$T$ limit $\hbar\omega/\kB T \ll 1$, the following limits hold: 
$c_2\to 1/\sqrt{2}$, $Q_{1,0}\to  \kB T \, / \, \hbar\omega = 2\pi (\aho/\LdB)^2$, and $z_{\nu} \simeq N_{\nu}\hbar\omega / \kB  T$. Finally, we find
\begin{equation}
\mathcal{C_{\nu}}=\frac{1}{\pi^{3/2} \aho^3}\sqrt{\frac{\kB T}{\hbar\omega}}N_{\nu}\sum_{\mu\ne\nu}N_{\mu}.
\end{equation}
In the peculiar case of a balanced mixture, we obtain
\begin{equation}
  \mathcal{C}_\mathrm{tot}=\frac{1}{\pi^{3/2} \aho^3}\sqrt{\frac{\kB T}{\hbar\omega}}\frac{\kappa-1}{\kappa}N^2,
  \label{contact-T}
\end{equation}
and if $\kappa\to\infty$, one recovers the result of~\cite{vignolo2013} for the Tonks-Girardeau gas.


\section{Details on the Matrix-Product-State (MPS) calculation}
\label{app:details-mps}

\textit{Method.---}
In order to determine density/momentum distributions and contacts in the finitely interacting model,
a variational two-site optimization-ansatz (DMRG) based on MPS was employed for ground-state finding~\cite{Schollwoeck2011}.
The code is based on the Abelian Symmetric Tensor Networks Library (developed in collaboration with the group of S. Montangero in Ulm),
which encompasses multiple Abelian symmetries (particle number conservation for each fermionic species).


\textit{Discretization of the problem.---}
To be able to do MPS-calculations on a discretized lattice, we transfer the original, continuous Hamiltonian into its tight-binding equivalent.
We therefore choose a region of the trap which is sufficiently large to contain the full ground-state ($\zeta a_\text{ho}$)
and cover it with a grid (in a-dimensional units, the region $y\in[-\zeta/2,+\zeta/2]$.
The lattice contains $L$ sites, i.e. we set $y_j = (j- (L+1)/2) \Delta y $ with $\Delta y = \zeta/L$ and  $j\in\{1,\ldots,L\}$ labeling the $j$-th site in the lattice. The discretized equivalent of the model in Eqs.~\eqref{eq:hamfree}-\eqref{eq:hamint} then reads
\begin{align}
\begin{split}
\label{eq:hamiltonian1}
 \hat{H} / \hbar \omega & =  -\frac{t}{2}\sum_{j=1}^{L-1}\sum_\sigma \left(c_{j,\sigma}^\dagger c_{j+1,\sigma} + \text{h.c.}\right)\\
 &\quad+ \sum_{j=1}^{L}\sum_\sigma \left(\frac{v}{2}\left(j-\frac{L+1}{2}\right)^2+t \right)c_{j,\sigma}^\dagger c_{j,\sigma}\\
 &\quad  + U \sum_{j=1}^{L} \sum_{\sigma \neq \sigma'} n_{j,\sigma} n_{j,\sigma'},
\end{split}
 \end{align}
where the index $j$ labels the lattice site and $\sigma$ labels the $\kappa$ internal $SU(\kappa)$-invariant levels (species).
The operators $c_{j,\sigma}$ and $c^\dagger_{j,\sigma}$ are the creation- resp. annihilation-operators for a fermion at site $j$ and of species $\sigma$,
and $n_{j,\sigma}$ measures the occupation of the site, $n_{j,\sigma}=c_{j,\sigma}^\dagger c_{j,\sigma}$.
The coefficients of this model relate to the ones of the dimensionless continuum model as $t = 1/\Delta y^2$, $v = \Delta y^2$ and $U= \alpha/\Delta y$.
Subsequently, the fermionic model is mapped to a spin-model.

All calculations were then performed with increasing numbers of sites (i.e. diminishing lattice spacing, $\Delta y \to 0$) 
and the quantities of interest were finite-size scaled to recover the continuum limit.

The maximum number of lattice sites considered was 216 ($\zeta \cong 12$ and $\Delta y \cong 0.055 a_\text{ho}$), and depending on the configuration, the virtual bond-dimension of the MPS was chosen $m=200-500$, corresponding to maximally discarded probabilities $\lesssim 10^{-6}$.

\textit{Observables.---} The Tan-contacts were calculated through measurement of the interaction-component in the ground-state energy. Density distributions were forthrightly determined through measurement of site-occupation in the discretized model; momentum distributions were obtained by Fourier-transformation of measured (fermionic) two-point correlators:
\begin{align}
n(k) = \frac{1}{L}\sum_{j,l=1}^{L} e^{i(j-l)k} \braket{c_j^\dagger c_l}.
\end{align}
In the numerical (spin) implementation, the fermionic two-point correlators are measured as a multi-point observable containing a Jordan-Wigner string.

\textit{Outlook.---} 
Incidentally, we notice that the identification of the proper Young tableau associated to the ground state thanks to the generalised Lieb-Mattis theorem
(explained above) would be of utmost utility as soon as our  MPS libraries will encompass non-Abelian symmetry groups as well (work in progress), implying both a huge speed-up and a large memory saving~\cite{Mila2014}. 

\begin{figure}
  \includegraphics[width=0.9\linewidth]{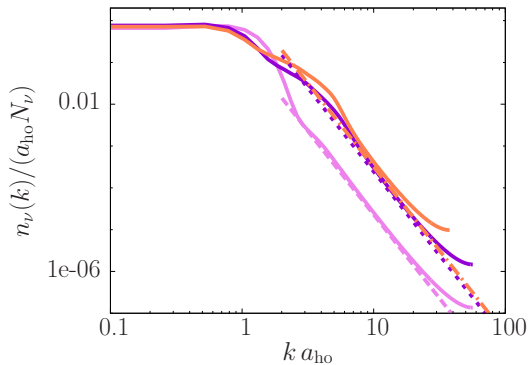}  
  \caption{\label{fig:momentum_tail}(Color online) 
  Numerical evidence for the existence of a generalized Tan's relation in a fermionic multi-component balanced mixture
  at finite interactions: $\gd=1\hbar\omega \aho$ (violet), $10\hbar\omega \aho$ (dark-violet), $100\hbar\omega \aho$ (orchide).
  As in Fig.~\ref{fig:mom}b-c, we display the case of $\kappa=3$ and $N=6$, 
  with the momentum distributions normalised to the species' population
  and plotted in units of $\aho$ versus the momentum in units of $1/\aho$.
  The solid curves are the momentum distributions $n_\nu(k)$ obtained by MPS numerics,,
  while the dashed straight lines are $\mathcal{C}_\nu k^{-4}$ where no free parameter is adjusted 
  and $\mathcal{C}_\nu$ are extracted from the interaction energies measured in the same simulations.
  } 
\end{figure}


\section{Details on the scaling results}
\label{app:details-scaling}
We start by rescaling the Hamiltonian $H = H_0 + H_\mathrm{int}$ of Eqs.(\ref{eq:hamfree}-\ref{eq:hamint})
in terms of the harmonic energy $\hbar \omega$ and length $\aho$, 
i.e., we define a new dimensionless coordinate $y_j = x_j / \aho$ and write:
\begin{equation}
\label{eq:hamscale}
H = \hbar \omega \left[
	\sum_{j=1}^{N} \left( -\frac{1}{2} \frac{\partial^2}{\partial y_j^2} + \frac{1}{2} y_j^2 \right) + \alpha \sum_{j<j^\prime} \delta (y_j - y_{j^\prime})
\right] \ , 
\end{equation}
where $\alpha = \aho m \gd / \hbar^2 = 2\aho/|a_{1\mathrm{D}}|$ is the natural choice for a  dimensionless parameter describing the relation between two typical lengths of the problem,
namely the 1D interaction  and the harmonic oscillator ones~\cite{Cazalilla2011}. 
The $O(N)$ orbitals occupied by the fermions give a contribution $O(N^2) \hbar \omega$ to the non-interacting energy,
and therefore make the energy per particle diverge in the thermodynamic limit (TL) $N\to \infty$,
unless we keep constant the chemical potential at $\alpha_0 \to \infty$, i.e., $\EF = N \hbar \omega$.
This corresponds to keeping the density $\rho = N / L$ fixed in the Bethe ansatz solution of the homogeneous system of length $L$~\cite{LiebLin,Yang67,Sutherland68,Takahashi70,LaiYang,Guan2012} ($\kF = \pi \rho$ and $\EF = \hbar^2 \kF^2 / 2m$ there).
The thermodynamic limit (TL) is then best studied then by introducing a further rescaling $y_j\to y_j/\sqrt{N}$
 \begin{equation} \label{eq:hamscale2}
H = \EF \left[
	\sum_{j=1}^{N} -\frac{1}{2} \frac{\partial^2}{\partial y_j^2} + \frac{1}{2 N^2} y_j^2 + \alpha_0 \sum_{j<j^\prime} \delta (y_j - y_{j^\prime})
\right] \, ,
\end{equation}
where the Hamiltonian in square brackets can now only give origin to a energy per particle which is a intensive dimensionless function of $\alpha_0$ only
(and, strictly speaking, of the single species polarization $p_\nu$, too):
\begin{equation}\label{eq:enscale-app}
	E / N = \EF f(\alpha_0, \{p_\nu\}) = N \, \hbar \omega \, f(\alpha / \sqrt{N}, \{p_\nu\}) ,
\end{equation}
as provided in the main text, Eq.~\eqref{eq:enscale}.
This can be easily verified in the extreme regimes $\alpha_0=0$ and $\alpha_0 \to \infty$, 
where respectively $f(0,\{p_\nu\}) = \sum_\nu p_\nu^2 / 2$ and $f(\infty,\{p_\nu\}) = 1/2$,
but remains at this stage a conjecture for intermediate regimes.
We remark, however, that the above corresponds also to taking the ratio between the interaction coefficient and the effective average density:
the latter can be estimated by the number of particles $N$ divided the \emph{rms} length of the highest occupied orbitals, which scales as $\sqrt{N}$.
This procedure is analogous to that encountered in the Bethe ansatz solution for the homogeneous system~\cite{LiebLin,Yang67,Sutherland68,Takahashi70,LaiYang,Guan2012} 
in order to write $E / N \propto \rho^2 e(\gamma)$ with $\gamma = m \gd / \hbar^2 \rho$,
which turns out to be valid at all values of $\gamma$.
Therefore, we have founded reasons to expect its validity in a large regime of $\alpha_0$.


\section{Details on the Bethe-Ansatz Local Density Approximation (BALDA)}
\label{app:details-balda}

\textit{General features.---}
Here we show the full derivation of the local density approximation to the density-functional approach theory,
under which the above conjectured scaling of Eq.~\eqref{eq:enscale-app} is formally exact for any $N$.
Practically, it is exact \emph{up to} corrections to LDA, indeed.
First, we can stay general about the composition of the mixture with total density $\rho(x)$, and define the energy functional 
\begin{equation}
E[\rho]=\int  [\epsilon[\rho,\gamma] + (V_\mathrm{ext}(x)-\mu)\rho(x)]dx\ ,
\label{functionalE}
\end{equation}
where $V_\mathrm{ext} =  m \omega^2 x^2 / 2$ and
\begin{equation}
\epsilon[\rho,\gamma]=\frac{\hbar^2}{2m} \rho^3 e(\gamma) \ ,
\label{epsilon}
\end{equation}
with $\gamma=m \gd /\hbar^2 \rho$ and $e(\gamma)$ is the dimensionless equation of state for the mixture  which can be obtained from Bethe Ansatz~\cite{LiebLin,Yang67,Sutherland68,Takahashi70,LaiYang,Guan2012}.
The density profile is then obtained by minimizing the energy ie $\delta E[\rho]/\delta \rho=0$
or equivalently $\delta \epsilon/\delta \rho=\mu-V_\mathrm{ext}$.
Using the above expression for $\epsilon$ we have
\begin{equation}
\frac{3}{2} \frac{\hbar^2}{m} \rho^2 e(\gamma) - \frac{\gd}{2} \rho e'(\gamma)=\mu -V_\mathrm{ext}(x)\ .
\label{rholda}
\end{equation}
The density is obtained by inversion of the above equation together with the normalization condition $\int \rho(x) dx=N$ which fixes the chemical potential,
and the positiveness of the r.h.s., which fixes a compact support $x \in [-R, R]$, with $R = \aho \sqrt{2 N \mu / \EF}$ the Thomas-Fermi radius. 

In order to obtain the contact we use the Hellmann-Feynman theorem $E_\mathrm{int}= \gd \partial E/\partial \gd$,
together with the Tan's relation~\eqref{eq:gtan2}, $\mathcal{C}_\mathrm{tot} = E_\mathrm{int} \gd m^2/ \pi\hbar^4$: 
\begin{equation}
\mathcal{C}_\mathrm{tot} = \frac{\gd^2 m^2}{2 \pi \hbar^4} \, \int dx \, \rho^2(x) \, e^\prime\left(\frac{m \gd}{\hbar^2 \rho(x)}\right).
\label{clda}
\end{equation}
The combined solution of Eqs.(\ref{clda}) and (\ref{rholda}) yields the LDA expression for the contact. 

\begin{figure}
\includegraphics[width=0.9\linewidth]{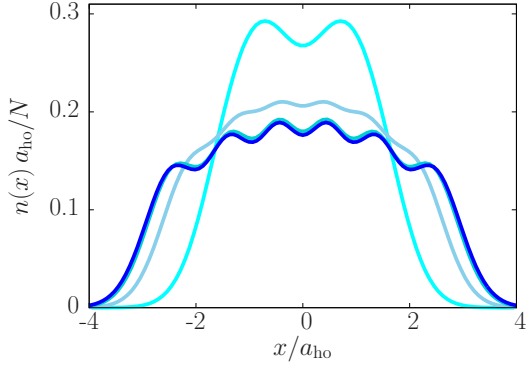}  
  \caption{  \label{fig:dens}(Color online)  Normalized density distributions $n(x)/N$ in units of $1/\aho$,  as a function of $x/\aho$, for the case of balanced mixtures with $N=6$ and $\kappa=3$. From the highest curve to the lowest: $\gd=0$, $\gd=1 \hbar\omega\aho$, $\gd=10 \hbar\omega\aho$, $\gd=100 \hbar\omega\aho$ and $\gd\to\infty$.}
\end{figure}

The scaling property~\eqref{eq:enscale-app} becomes apparent by the following change of variables in Eq.~\eqref{rholda}:
$\mu \equiv \tilde{\mu} \EF$, $z = x/R$ and $r(z) = \aho \rho(x) / \sqrt{N}$ (where we conveniently leave out a factor $\sqrt{2\tilde{\mu}}$),
giving
\begin{equation}
\label{eq:rholda1}
3 r(z)^2 e\left(\frac{\alpha_0}{r(z)}\right) - \alpha_0 r(z) e^\prime\left(\frac{\alpha_0}{r(z)}\right) = 2\tilde{\mu} (1-z^2) ,
\end{equation}
with normalization condition $\sqrt{2\tilde{\mu}} \int_{-1}^{1} r(z) \mathrm{d}z  = 1$.
It is indeed immediate to see that any dependence on $N$ drops out from the solution,
apart from the factors involved in the change of variables,
and that everything will be a function of $\alpha_0$ only.
The total energy can be computed by formally integrating $E =  \hbar \omega \, \int N \tilde{\mu}(\alpha / \sqrt{N}) \, \mathrm{d}N$,
which by taking $\alpha_0$ as integration variable is equivalent to:
\begin{equation}\label{eq:enscaleBALDA}
	f_{\mathrm{BALDA}}(\alpha_0, \{p_\nu\}) \equiv \alpha_0^4 \int \frac{-2 \tilde{\mu}(\alpha_0)}{\alpha_0^5} \mathrm{d}\alpha_0 ,
\end{equation}
and this concludes our proof.

\textit{Balanced mixture.---}
For the strong coupling regime $\alpha_0 \to \infty$, in the specific case of a balanced mixture
we can resort to a series' expansion of the dimensionless equation of state~\cite{Guan2012},
\begin{align}\label{egamma}
\begin{split}
e(\gamma) = \frac{\pi^2}{3} & \left[
	1 - \frac{4 Z_1(\kappa)}{\gamma} + \frac{12 Z_1(\kappa)^2}{\gamma^2} \right. \\
	 & \left.	- \frac{32}{\gamma^3} \left( Z_1(\kappa)^3 - \frac{Z_3(\kappa) \, \pi^2}{15} \right)
		+O\left(\frac{1}{\gamma^4}\right)
\right] 
\end{split}
 \end{align}
(with $Z_1$ and $Z_3$ defined in the main text), and similarly expand $\tilde{\mu}$ and $r(z)$ as function of $1/\alpha_0$ to solve Eq.~\eqref{eq:rholda1}~\cite{notebalda}.
By then plugging the results into Eq.~\eqref{clda}, we finally obtain the perturbative expressions 
for the contacts~\eqref{eq:CbaldaINF}-\eqref{eq:Cbaldafiniteg}.
We remark here that the leading term~\eqref{eq:CbaldaINF} can be readily obtained 
by employing the well-known analytical LDA profile at infinite interactions, $\rho(x)=\sqrt{2N} \sqrt{1 - (x/R)^2} / \pi \aho$,
and the lowest non-zero order of the derivative $e^\prime \simeq 4 \pi^2 Z_1(\kappa) / 3 \gamma^2$ inside Eq.~\eqref{clda}.

Finally, we stress that strong coupling expansions have also been provided in Ref.~\cite{Guan2012} for imbalanced mixtures, and they can be plugged into the above procedure to extract the expressions for the contacts of each species for the specific experimental setup.


\section{Construction of the Young tableaux}
\label{const-young}

Here we specify how to determine which Young tableaux are possible given a specific Fermi mixture, and provide some examples in the case of $N=6$ fermions.
We recall that, in a Young tableau, boxes belonging to the same line (resp. column) correspond to a symmetric (resp. antisymmetric) exchange of particles \cite{Hamermesh_book}. The only restriction is that two fermions belonging to the same component must follow the Pauli principle.
Trivially, for a single-component mixture, the only possible tableau is {\tiny\Yvcentermath1$\yng(1,1,1,1,1,1)$}.
In non-trivial cases, the following procedure is employed: first, we label the species by a letter, starting from $a$ for the most populated species and ordering them by decreasing order of population.
We then build the tableaux by labeling their boxes with the previous species labels, imposing that each label must appear only one time per row.
For example, in the case of a two-component balanced mixture, the possibilities are : {\tiny\Yvcentermath1$\young(ab,ab,ab)$}, {\tiny\Yvcentermath1$\young(ab,ab,a,b)$}, {\tiny\Yvcentermath1$\young(ab,a,a,b,b)$} and {\tiny\Yvcentermath1$\young(a,a,a,b,b,b)$}.
Note that when summing the dimensions of these tableaux (which are given by the hook-length formula \cite{Hamermesh_book}), one recovers exactly the dimension of the degenerate manifold : in the previous example $5+9+5+1=20=\frac{6!}{3!3!}$.


\section{Density profiles at finite interactions}
\label{details-nx}
We illustrate here the density profiles of the multicomponent mixtures at various interaction strengths, as compared with the density profiles obtained from the exact solution. As shown in Fig. \ref{fig:dens}, at reduced interaction strength $\alpha_0= 41$ ($\gd=100\hbar\omega\aho$), the profiles are almost indistinguishable from the ones of the exact solution at infinite interactions.

\end{document}